\documentstyle[prd,aps]{revtex}
\input epsf

\begin{document}
\draft
\twocolumn[\hsize\textwidth\columnwidth\hsize\csname
@twocolumnfalse\endcsname
\preprint{PACS: 98.80.Cq, CERN-TH/99-77,\\
{}~hep-ph/yymmnn}

\def\simlt{\stackrel{<}{{}_\sim}}
\def\simgt{\stackrel{>}{{}_\sim}}
\newcommand\lsim{\mathrel{\rlap{\lower4pt\hbox{\hskip1pt$\sim$}}
    \raise1pt\hbox{$<$}}}
\newcommand\gsim{\mathrel{\rlap{\lower4pt\hbox{\hskip1pt$\sim$}}
    \raise1pt\hbox{$>$}}}

\title{Measuring the Cosmological Lepton Asymmetry through the CMB Anisotropy}

\author{William H. Kinney$^{(1)}$ and   Antonio
Riotto$^{(2)}$}

\address{$^{(1)}${\it Univ. of Florida Dept. of Physics, P.O. Box 118440, Gainesville, FL 32611}}

\address{$^{(2)}${\it CERN, Theory Division, CH-1211 Geneva 23, Switzerland}}

\date{\today}
\maketitle

\begin{abstract}

A large lepton asymmetry in the Universe is still  a viable possibility and leads to many interesting phenomena such as gauge symmetry 
nonrestoration at high temperature. We show that 
a large lepton asymmetry  changes  
 the predicted cosmic microwave background (CMB) anisotropy and that any degeneracy in the relic neutrino sea will be measured to a precision of 1\% or better when  the CMB anisotropy is measured at the accuracy expected to result from the planned satellite missions  MAP and Planck. In fact, the current measurements already put an upper limit on the lepton asymmetry of the Universe which is stronger than the one coming from considerations of primordial nucleosynthesis and structure formation.

\end{abstract}

\pacs{PACS: 98.80.Cq    \hskip 1 cm CERN-TH/99-77 \hskip 1 cm UFIFT-HEP-99-3}
\vskip1pc]

{\it 1. Introduction}~~ For all we know,   the Universe may contain  background charges which are comparable to, if not larger than, the entropy density. The presence of some 
sizeable charge asymmetry may postpone symmetry restoration 
in nonsupersymmetric theories  \cite{hw82} or -- even more remarkably -- 
it can lead to symmetry breaking of internal symmetries at high 
temperature $T$ \cite{bbd91}. Furthermore, the phenomenon 
of symmetry nonrestoration at high $T$ in presence of 
large charge asymmetries  has been 
recently shown to work in supersymmetry too \cite{rs97}. 

The principal candidate for a large charge is the  lepton number  which 
today could reside in the form of 
neutrinos. This has inspired Linde in his original work to point out that 
large enough lepton number of the Universe would imply the nonrestoration 
of symmetry even in the Standard  Model (SM) \cite{l79}. The fact 
that the large lepton number can be consistent with the small baryon 
number asymmetry \cite{reviewbau} in the context of grand unification has been pointed out 
a long time ago \cite{hk81} and recently a model for producing large 
$L$ and small $B$ has been presented \cite{ccg97}.
Moreover, while one could 
naively think that the large lepton number 
would be washed out by the sphaleron effects
at the temperature above the weak scale, it turns out  that the
nonrestoration of symmetry prevents this from happening \cite{ls94}.

Remarkably 
enough, having a large lepton asymmetry   still remains a consistent possibility.
The successful predictions of primordial nucleosynthesis are 
not jeopardized as long as the neutrino degeneracy parameter $\xi_\nu=\mu_\nu/T$, where $\mu_\nu$ is the chemical potential of the degenerate neutrinos, is small enough. 
Combining the nucleosynthesis bounds \cite{ks92} with the ones coming from structure formation in the Universe \cite{bound} yields $-0.06 {\mathrel{\rlap{\lower4pt\hbox{\hskip1pt$\sim$}}
    \raise1pt\hbox{$<$}}}\xi_{\nu_e} {\mathrel{\rlap{\lower4pt\hbox{\hskip1pt$\sim$}}
    \raise1pt\hbox{$<$}}} 1.1$ and $\left|\xi_{\nu_{\mu,\tau}}\right|{\mathrel{\rlap{\lower4pt\hbox{\hskip1pt$\sim$}}\raise1pt\hbox{$<$}}} 6.9$. 

It is quite intriguing that the very simple and economical  hypothesis of large  degeneracy  in the sea of relic neutrinos may lead to so many  interesting phenomena in the early Universe.
If Nature has chosen the  option that the lepton number 
is large enough so that SM symmetry (or extensions of it) is not restored at high temperature --
but without any charge field 
condensation --    the cosmological consequence 
would be remarkable, for this would suffice to nonrestore the symmetry. Symmetry  nonrestoration provides   a simple way out of the 
the monopole problem and the domain wall problem   \cite{bsr1,bsr2} which are some of the central issues in the modern astroparticle physics and are 
 especially serious being  generic to the idea
of grand unification. 
Thus, if the lepton number of the Universe were to turn 
out large, there would be no monopole and domain problems whatsoever. A neutrino degeneracy $\xi_\nu$ at temperatures above 100 GeV in the range  $(2.5-5.3)$ for the SM Higgs boson mass in the interval $(100-800)$ GeV would suffice
to avoid the SM gauge symmetry restoration in the hot Universe \cite{bsr1}.   

Moreover, if the lepton number density of the Universe is of order of the entropy density, the neutrinos with masses in the Super-Kamiokande range $\sim 0.07$ eV \cite{atmospheric} can make a significant contribution to the energy density of the Universe \cite{pal} or even explain the cosmic rays with energies in excess of the Greisen-Zatsepin-Kuzmin cutoff \cite{ray}. This would require a value of the neutrino degeneracy parameter of the order of 4.6.

The main point we wish to make, however, is that the most  spectacular cosmological consequence of a large lepton asymmetry in the Universe is  its impact  on the temperature variations of the CMB radiation. 

The CMB  provides a  window on fundamental physics at very high energy scales \cite{review} and
the measurement of the  spectral index $n$, specifying the scale-dependence of the spectrum of the curvature perturbation, will be a powerful discriminator between models of inflation \cite{lr}, when it is measured at the accuracy expected to result from the planned satellite missions  MAP and Planck  \cite{MAP+PLANCK}. Furthermore,  
observations of the polarization of the CMB have the potential to place much tighter constraints on cosmological parameters than observations of the fluctuations in temperature alone. The detection of  a tensor/scalar ratio $r \sim 0.01$ would allow precision tests of interesting inflation models, and is possible with a modest increase in sensitivity over that planned for the Planck satellite, or potentially by ground-based experiments \cite{pol}.

In this Letter we show that  a large lepton asymmetry in the Universe leads to
drastic changes in the predicted CMB anisotropies that might be unambiguously detected by future satellite experiments. This will allow us   to test the  presence of neutrino chemical potentials $(\mu_\nu/ T)$ to a precision of 1\%  or better. The   precision increases considerably with the value of the neutrino degeneracy. This is exactly the situation one would hope for since most of the current speculations   make use of  large neutrino degeneracies. This, in turn, will give us an enormous amount of information about the dynamical evolution of the early Universe.   Many intriguing ideas such as  the possibility that the  some gauge symmetries  were never restored in the hot Universe because of a large lepton charge -- will be tested. 

In fact, the current information on the CMB anisotropy already places an upper limit on the lepton asymmetry of the Universe  stronger than other considerations. We will argue that the present data  disfavour values of the neutrino chemical potential $(\mu_\nu / T)$ larger than about 5, independent of the neutrino flavour.  The present level of knowledge about the CMB fluctuation spectrum is not only sufficient to place  bounds which are more severe than the ones coming from considerations about nucleosynthesis and structure formation, but also to put meaningful bounds on some speculations about the evolution of the early Universe.

{\it 2. Lepton asymmetry and present CMB data}~~An antisymmetry between neutrinos and antineutrinos in the universe is most conveniently measured by the chemical potential $\mu_\nu$ between the two species. The difference in neutrino number density $n_{\nu}$ and the antineutrino number density $n_{\bar\nu}$ for a single degenerate neutrino species can be expressed as
\begin{eqnarray}
n_{\nu} - n_{\bar\nu} =&& {T^3 \over 2 \pi^2} \int_{m_\nu}^{\infty}{u\,du \sqrt{u^2 - m^2}}\cr
&&\times \left({1 \over 1 + \exp\left(u - \xi_\nu\right)} - {1 \over 1 + \exp\left(u + \xi_\nu\right)}\right),
\end{eqnarray}
where $u \equiv E_\nu / T$, and $\xi_\nu \equiv \mu_\nu / T$. If the neutrinos are relativistic, $m_{\nu} \ll T$, the cosmological lepton asymmetry can be written as the ratio of the neutrino asymmetry to the entropy $
L \equiv (n_{\nu} - n_{\bar \nu})/s = {15 \over 4 \pi^4 g_{*S}} \left(T_{\nu}/ T_{\gamma}\right)^3 \left(\pi^2 \xi_\nu + \xi_\nu^3\right)$.
The lepton asymmetry $L$ is conserved in the cosmological expansion, and, as long as the neutrinos remain relativistic after, so is $\xi$. Even relatively heavy neutrinos will be relativistic until well after recombination, so for the purposes of investigating effects on the CMB, we can safely take $\xi$ as constant.  Yet, the reader should bear in mind  that at temperatures larger than $\sim 1$ MeV $\xi_\nu\propto g^{1/3}_{*S}$ if $\xi_\nu$ is larger than unity and that our bounds  refer to the present values of $\xi_\nu$.

We will assume   that the lepton asymmetry in the neutrino sector occurs in second family ($\mu$ neutrinos) or third family ($\tau$ neutrinos), so that direct effects on primordial nucleosynthesis are   absent.
If both neutrino families carry a chemical potential the effect on the CMB is enhanced over that for a single species. 
 The effect of the neutrino degeneracy is then confined to: {\it i)} changing the time of matter/radiation equality, and {\it ii)} changing the time of neutrino decoupling\cite{ks92}. We will further assume that the neutrinos are light enough to remain relativistic until well after recombination. This is a good approximation for neutrinos with masses in the Super-Kamiokande range. 

The evaluation of the effect of the neutrino degeneracy on the CMB requires numerical evaluation of a Boltzmann equation. We use Seljak and Zaldarriaga's CMBFAST code \cite{seljak96} to calculate the CMB multipole spectrum, described in detail in the next section. The results are presented in Fig. 1 which shows the CMB spectrum for various values of the lepton asymmetry for a {\it single} family, along with the results of current CMB experiments as compiled by Tegmark\cite{tegwebsite}. Even though the current data are not conclusive in placing strict limits on the cosmological lepton asymmetry, it is evident that values of $\xi_\nu$ larger than about 5 are a poor fit to the existing data. 

\begin{figure}
\centerline {\epsfysize=3.0in \epsfbox{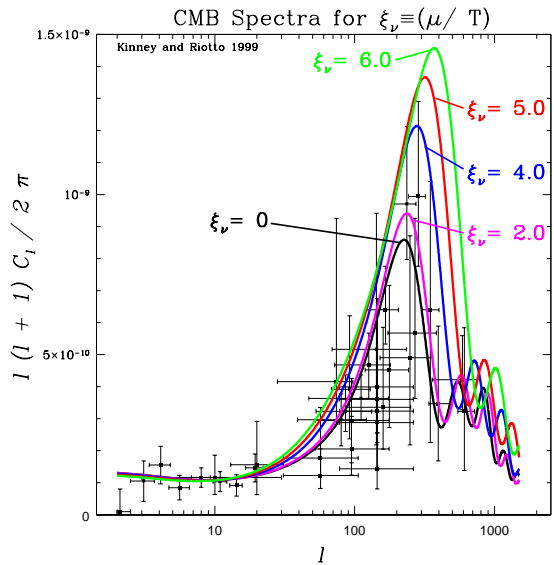} }
\caption{CMB spectra for various $\xi_\nu \equiv (\mu_\nu / T)$, for a background cosmology with $h = 0.65$, $\Omega_M = 0.3$ and $\Omega_\Lambda = 0.7$. Points with error bars are currently available CMB data.}
\label{figscalarspectra}
\end{figure}

Future experiments are likely to tighten the error bars significantly. In the next section, we discuss the CMB spectrum in detail, and discuss the prospects for future experiments, particularly NASA's MAP satellite and the ESA's Planck Surveyor, to place limits on the cosmological lepton asymmetry.

{\it 3. Statistics of CMB measurements: temperature and polarization}~~
What observations of the cosmic microwave background actually measure is anisotropy in the temperature of the radiation as a function of direction. It is natural to expand the anisotropy on the sky in spherical harmonics:
\begin{equation}
{\delta T\left(\theta,\phi\right) \over T_0} = \sum_{l = 0}^{\infty}\sum_{m = -l}^{l}{a^T_{lm} Y_{lm}\left(\theta,\phi\right)},
\end{equation}
where $T_0 = 2.728^\circ K$ is the mean temperature of the CMB. Inflation predicts that each $a^T_{lm}$ will be Gaussian distributed with mean $\left\langle a^T_{lm} \right\rangle = 0$ and variance $
\left\langle a^{T*}_{l'm'} a^T_{lm}\right\rangle = C_{Tl} \delta_{ll'} \delta_{mm'}$, 
where angle brackets indicate an average over realizations. For Gaussian fluctuations, the set of $C_{Tl}$'s completely characterizes the temperature anisotropy. The spectrum of the $C_{Tl}$'s is in turn dependent on cosmological parameters such as $\Omega_0$, $H_0$, $\Omega_{\rm B}$ and so forth, so that observation of CMB temperature anisotropy can serve as an exquisitely precise probe of cosmological models. 

The cosmic microwave background is also expected to be polarized due to the presence of fluctuations. Observation of polarization in the CMB will greatly increase the amount of information available for use in constraining cosmological models. Polarization is a {\it tensor} quantity, which can be decomposed on the celestial sphere into ``electric-type'', or scalar, and ``magnetic-type'', or pseudoscalar modes. The symmetric, trace-free polarization tensor ${\cal P}_{ab}$ can be expanded as\cite{kamionkowski96}
\begin{equation}
{{\cal P}_{ab} \over T_0} =  \sum_{l = 0}^{\infty}\sum_{m = -l}^{l} \left[a^E_{lm} Y^E_{\left(l m\right) ab}\left(\theta,\phi\right) + a^B_{lm} Y^B_{\left(l m\right) ab}\left(\theta,\phi\right)\right],
\end{equation}
where the $Y^{E,B}_{\left(l m\right) a b}$ are electric- and magnetic-type tensor spherical harmonics, with parity $(-1)^l$ and $(-1)^{l + 1}$, respectively. Unlike a temperature-only map, which is described by the single multipole spectrum of $C^T_l$'s, a temperature/polarization map is described by three spectra
\begin{equation}
\left\langle \left|a^T_{lm}\right|^2\right\rangle \equiv C_{Tl},\ \left\langle\left|a^E_{lm}\right|^2\right\rangle \equiv C_{El},\ \left\langle \left|a^B_{lm}\right|^2\right\rangle \equiv C_{Bl},
\end{equation}
and three correlation functions,
$\left\langle a^{T*}_{lm} a^E_{lm}\right\rangle \equiv C_{Cl}$, $\left\langle a^{T*}_{lm} a^B_{lm} \right\rangle \equiv C_{(TB)l}$,
$\left\langle a^{E*}_{lm} a^B_{lm}\right\rangle \equiv C_{(EB)l}$.
Parity requires that the last two correlation functions vanish, $C_{(TB)l} = C_{(EB)l} = 0$, leaving four spectra: temperature $C_{Tl}$, E-mode $C_{El}$, B-mode $C_{Bl}$, and the cross-correlation $C_{Cl}$. Since scalar density perturbations have no ``handedness,'' it is impossible for scalar modes to produce B-mode (pseudoscalar) polarization. Only tensor fluctuations (or foregrounds \cite{zaldarriaga98}) can produce a B-mode. 

Measurement uncertainty in cosmological parameters is characterized by the Fisher information matrix $\alpha_{ij}$. (For a review, see Ref. \cite{tegmark97}.) Given a set of parameters $\left\lbrace \lambda_i \right\rbrace$, the Fisher matrix is given by
\begin{equation}
\alpha_{ij} = \sum_l \sum_{X,Y} {\partial C_{Xl} \over \partial \lambda_i} {\rm Cov}^{-1}\left(\hat C_{Xl} \hat C_{Yl}\right)  {\partial C_{Yl} \over \partial \lambda_j},
\end{equation}
where $X,Y = T,E,B,C$ and $\rm{Cov}^{-1}\left(\hat C_{Xl} \hat C_{Yl}\right)$ is the inverse of the covariance matrix between the estimators $\hat C_{Xl}$ of the power spectra. Calculation of the Fisher matrix requires assuming a ``true'' set of parameters and numerically evaluating the $C_{Xl}$'s and their derivatives relative to that parameter choice. The covariance matrix for the parameters $\left\lbrace \lambda_i\right\rbrace$ is just the inverse of the Fisher matrix, $\left(\alpha^{-1}\right)_{ij}$, and the expected error in the parameter $\lambda_i$ is of order $\sqrt{\left(\alpha^{-1}\right)_{ii}}$. The full set of parameters $\left\lbrace \lambda_i \right\rbrace$ we allow to vary is: 1)the tensor/scalar ratio $r$, 2) the spectral index $n$, 3) the normalization $Q_{\rm rms-PS}$, 4)  the baryon density $\Omega_{\rm B}$, 5)
the Hubble constant $h \equiv H_0 / (100\,{\rm km\,sec^{-1}\,Mpc^{-1}})$, 6) 
the reionization optical depth, $\tau_{\rm ri}$ and 7) the 
 the neutrino chemical potential $\left(\mu_\nu / T\right)$.

We take as a ``fiducial'' model COBE normalization \cite{bunn96} with $\Omega_{\rm B} = 0.05$ and $h = 0.65$, and a reionization optical depth of $\tau_{\rm ri} = 0.05$, corresponding to reionization at a redshift of about $z \sim 13$. The tensor/scalar ratio is $r = 0$. Fixed parameters are $\Omega_M = 0.3$ and $\Omega_\Lambda = 0.7$, consistent with inflation. 
Assuming an approximately gaussian beam, the nonzero elements of the covariance matrix ${\rm Cov}\left(\hat C_{Xl} \hat C_{Yl}\right)$ are \cite{knox95,seljak96a,zaldarriaga97b,kamionkowski96}
\begin{eqnarray}
{\rm Cov}\left(\hat C_{Tl} \hat C_{Tl}\right) =&& {2 \over \left(2 l + 1\right) f_{\rm sky}} \left(C_{Tl} + w^{-1}_T e^{l^2 \sigma^2_{\rm b}}\right)^2,\cr
{\rm Cov}\left(\hat C_{El} \hat C_{El}\right) =&& {2 \over \left(2 l + 1\right) f_{\rm sky}} \left(C_{El} + w^{-1}_P e^{l^2 \sigma^2_{\rm b}}\right)^2,\cr
{\rm Cov}\left(\hat C_{Bl} \hat C_{Bl}\right) =&& {2 \over \left(2 l + 1\right) f_{\rm sky}} \left(C_{Bl} + w^{-1}_P e^{l^2 \sigma^2_{\rm b}} \right)^2,\cr
{\rm Cov}\left(\hat C_{Cl} \hat C_{Cl}\right) =&& {2 \over \left(2 l + 1\right) f_{\rm sky}} \left[C^2_{Cl} +\right.\cr
&&\left.\left(C_{Tl} + w^{-1}_T e^{l^2 \sigma^2_{\rm b}}\right) \times\right.\cr
&&\left.\left(C_{El} + w^{-1}_P e^{l^2 \sigma^2_{\rm b}}\right)\right],\cr
{\rm Cov}\left(\hat C_{Tl} \hat C_{El}\right) =&& {2 \over \left(2 l + 1\right) f_{\rm sky}} C^2_{Cl},\cr
{\rm Cov}\left(\hat C_{Tl} \hat C_{Cl}\right) =&& {2 \over \left(2 l + 1\right) f_{\rm sky}} C_{Cl} \left(C_{Tl} + w^{-1}_T e^{l^2 \sigma^2_{\rm b}} \right),\cr
{\rm Cov}\left(\hat C_{El} \hat C_{Cl}\right) =&& {2 \over \left(2 l + 1\right) f_{\rm sky}}  C_{Cl} \left(C_{El} + w^{-1}_P e^{l^2 \sigma^2_{\rm b}} \right).
\end{eqnarray}
Here $f_{\rm sky}$ is the fraction of the sky observed, and $\sigma_{\rm b} = \theta_{\rm fwhm} / \sqrt{8 \ln 2}$ is the gaussian beamwidth, where $\theta_{\rm fwhm}$ is the full width at half maximum. The inverse weights per unit area $w^{-1}_T$ and $w^{-1}_P$ are determined by the detector resolution and sensitivity. For a noise per pixel $\sigma^T_{\rm pixel}$ and solid angle per pixel $\Omega_{\rm pixel} \simeq \theta_{\rm fwhm}^2$, the weight $w^{-1}_T$ is
\begin{equation}
w^{-1}_T = {\sigma^2_{\rm pixel} \Omega_{\rm pixel} \over T_0^2}.
\end{equation}
The polarization pixel noise $\sigma^P_{\rm pixel}$ is simply related to the temperature pixel noise $\sigma^T_{\rm pixel}$, since the number of photons available for the temperature measurement is twice that for the polarization measurements: $
\left(\sigma^P_{\rm pixel}\right)^2 = 2 \left(\sigma^T_{\rm pixel}\right)^2$
and $w^{-1}_P = 2 w^{-1}_T$. For an observation with multiple channels $c$ with
 different beam sizes and sensitivities, the weights $w_T^{\left(c\right)}$
simply add\cite{bond97}. For MAP, we combine the three high-frequency channels at $40$, $60$, and $90\,{\rm GHz}$, each with a pixel noise of $\sigma_{\rm pixel} = 35\,{\rm \mu K}$ and beam sizes $\theta_{\rm fwhm} = (28.2',21.0',12.6')$ respectively. Similarly, for Planck we combine the two channels at $143$ and $217\,\rm{GHz}$, with beam width $\theta_{\rm fwhm} = (8.0',5.5')$ and  pixel noise $\sigma^T_{\rm pixel} = (5.5\,{\rm \mu K},11.7\,{\rm \mu K})$. In all cases we take the sky fraction to be $f_{\rm sky} = 0.65$. The Table  shows the expected measurement uncertainty at the $1\sigma$ level in $\xi_\nu$ for various values of the lepton asymmetry.

\begin{center}
\begin{tabular}{|c|c|c|}
\hline
$\xi_\nu $ & $(\delta\xi_\nu)_{{\rm MAP}}$ & $(\delta\xi_\nu)_{{\rm Planck}}$ \\ \hline
0.25 & $--$ & 0.10  \\ \hline
0.5 & 0.4 & 0.05\\ \hline
1.0 & 0.2 & 0.02  \\ \hline
2.0 & 0.09 & 0.01  \\ \hline
4.0 & 0.04 & 0.005\\ 
\hline
\end{tabular}
\end{center}

We see that the expected measurement errors drop sharply as $\xi$ increases, with measurement errors of order a percent possible for large $\xi$. MAP is capable of a marginal detection of $\xi = 0.5$, while Planck can detect $\xi$ a factor of two smaller. What is exciting is that the uncertainties drop significantly for large values of $\xi_\nu$, which many of the speculative proposals make use of. 

{\it 4. Conclusions}~~
The lepton asymmetry of the universe is, at present, not a well-constrained quantity. In this Letter, we have shown that the cosmic microwave background is a  powerful tool for placing constraints on the cosmological lepton asymmetry. In fact, the current state of knowledge about the CMB spectrum already allows useful conclusions to be drawn. (For an analysis of current constraints, see Ref. \cite{lesgourgues99}.)
In the light of our results we may argue that  the solution to the  monopole problem in Grand Unified Theories as well to  the domain wall problem by storing a large  lepton number asymmetry in the Universe is starting to be challenged by present CMB data. The same conclusion may be drawn for the suggestion that the ultra-high energy cosmic rays  beyond the Greisen-Zatsepin-Kuzmin cutoff may be explained with the aid of a  neutrino degeneracy of $\sim 4.6$ \cite{ray}. 
 
But -- luckily --  the best is still to come. 
Future satellite experiments  promise to greatly improve our knowledge of the lepton asymmetry of the Universe, with uncertainty in the chemical potential of a degenerate neutrino species of order a percent or better within reach of planned experiments. It is  intriguing that future measurements of the CMB anisotropy  at the expected accuracy can tell us so much about the early evolution of the Universe. 

We would like to thank Julien Lesgourgues for helpful correspondence. WHK is supported by U.S. DOE grant DEFG05-86ER-40272.


\begin{thebibliography}{10}

\bibitem{hw82}
H.~E. Haber and H.~A. Weldon, Phys. Rev. {\bf D25},  502  (1982).

\bibitem{bbd91}
K.~M. Benson, J. Bernstein, and S. Dodelson, Phys. Rev. {\bf D44},  2480
  (1991).

\bibitem{rs97}
A. Riotto and G. Senjanovi\' c, Phys. Rev. Lett. {\bf 79},  349  (1997), hep-ph/9702319.



\bibitem{l79}
A.~D. Linde, Phys. Lett. {\bf 86B},  39  (1979).


\bibitem{reviewbau} For a recent review, see A. Riotto and M. Trodden,  hep-ph/9901362, to appear in An. Rev. of Nucl. and Part. Sci.


\bibitem{hk81}
J.~A. Harvey and E.~W. Kolb, Phys. Rev. {\bf D24},  2090  (1981).

\bibitem{ccg97}
A. Casas, W.-Y. Cheng, and G. Gelmini, Nucl. Phys. {\bf 538}, 297 (1999).


\bibitem{ls94}
J. Liu and G. Segre, Phys. Lett. {\bf B338},  259  (1994).

\bibitem{ks92}
H.-S. Kang and G. Steigman, Nucl. Phys. {\bf B372},  494  (1992), 
and references therein.

\bibitem{bound} K. Freese, E.W. Kolb and M.S. Turner, Phys. Rev. {\bf D27}, 1689 (1983).

\bibitem{bsr1} B. Bajc, A. Riotto and  G. Senjanovi\'c, Phys. Rev. Lett. {\bf 81}, 1355 (1998), hep-ph/9710415. 

\bibitem{bsr2} B. Bajc, A. Riotto and  G. Senjanovi\'c, Mod. Phys. Lett. {\bf A13}, 2955 (1998), hep-ph/9803438.

\bibitem{atmospheric}
  Y.~Fukuda et al. (Super-Kamiokande Collaboration), hep-ex/9807003.


\bibitem{pal} P. Pal and K. Kar, hep-ph/9809410.

\bibitem{ray} G. Gelmini and A. Kusenko, hep-ph/9902354.




\bibitem{review} For a short review, see J. Ellis, astro-ph/9902242.

\bibitem{lr} D.H. Lyth and A. Riotto, hep-ph/9807278, to be published in Phys. Rept. 

\bibitem{MAP+PLANCK}
  See http://map.gsfc.nasa.gov for information on MAP and
  http://astro.estec.esa.nl/Planck/ for Planck.
  

\bibitem{pol} W.H. Kinney, Phys. Rev. {\bf D58}, 123506 (1998).

\bibitem{seljak96} U. Seljak and M. Zaldarriaga, Astrophys. J. {\bf 469}, 437 (1996).

\bibitem{tegwebsite} See http://www.sns.ias.edu/~max/cmb/experiments.html.

\bibitem{kamionkowski96} M. Kamionkowski, A. Kosowsky, and A. Stebbins, Phys. Rev. D {\bf 55}, 7368 (1997).


\bibitem{kamionkowski96} M. Kamionkowski, A. Kosowsky, and A. Stebbins, Phys. Rev. D {\bf 55}, 7368 (1997).

\bibitem{zaldarriaga98} M. Zaldarriaga and U. Seljak, Report No. astro-ph/9803150.

\bibitem{tegmark97} M. Tegmark, A. Taylor, and A. Heavens, Astrophys. J. {\bf 480} 22 (1997).

\bibitem{bunn96} E. F. Bunn and M. White, Astrophys. J. {\bf 480}, 6 (1997).

\bibitem{knox95} L. Knox, Phys. Rev. D {\bf 52}, 4307 (1995).

\bibitem{seljak96a} U. Seljak, Astrophys. J. {\bf 482}, 6 (1996).

\bibitem{zaldarriaga97b} M. Zaldarriaga and U. Seljak, Phys. Rev. D {\bf 55} 1830 (1997).

\bibitem{bond97} J. R. Bond, G. Efstathiou, and M. Tegmark, Mon. Not. R. Astron. Soc. {\bf 291}, L33 (1997).

\bibitem{lesgourgues99} J. Lesgourgues and S. Pastor, hep-ph/9904411.


\end{thebibliography}
\end{document}